\pdfoutput=1

\documentclass[10pt,letterpaper]{article}
\usepackage[top=0.85in,left=2.75in,footskip=0.75in]{geometry}

\usepackage{amsmath,amssymb}

\usepackage{changepage}

\usepackage[utf8x]{inputenc}

\usepackage{textcomp,marvosym}

\usepackage{cite}

\usepackage{nameref,hyperref}

\usepackage[right]{lineno}

\usepackage{microtype}
\DisableLigatures[f]{encoding = *, family = * }

\usepackage[table]{xcolor}

\usepackage{array}

\newcolumntype{+}{!{\vrule width 2pt}}

\newlength\savedwidth

\raggedright
\usepackage[aboveskip=1pt,labelfont=bf,labelsep=period,justification=raggedright,singlelinecheck=off]{caption}

\makeatletter
\renewcommand{\@biblabel}[1]{\quad#1.}
\makeatother

\usepackage{lastpage,fancyhdr,graphicx}
\usepackage{epstopdf}
\fancyheadoffset[L]{2.25in}
\fancyfootoffset[L]{2.25in}
\usepackage{comment}
\usepackage[final]{pdfpages}

\begin{document}
\vspace*{0.2in}

\begin{flushleft}
{\Large
\textbf\newline{Enhanced precision of circadian rhythm by output system} 
}
\newline

Hotaka Kaji\textsuperscript{1},
Fumito Mori\textsuperscript{1,2},
Hiroshi Ito\textsuperscript{1*},

\bigskip
\textbf{1} Faculty of Design, Kyushu University, Fukuoka, Japan
\\
\textbf{2} Education and Research Center for Mathematical and Data Science, Kyushu University, Fukuoka, Japan
\\

\bigskip

* hito@design.kyushu-u.ac.jp
\end{flushleft}

\abstract{Circadian rhythms are biological rhythms of approximately 24 h that persist even under constant conditions without environmental daily cues. The molecular circadian clock machinery generates the physiological rhythms, which can be transmitted into the downstream output system. Owing to the stochastic nature of the biochemical reactions, the oscillation period of circadian rhythms exhibited by individual organisms or cells is not constant on a daily basis with variations as high as 10 \% in terms of the coefficient of variation. Although the fluctuations in circadian rhythm is measured through a reporter such as bioluminescence or fluorescence, experimentally confirming whether the fluctuations found in the reporter system are the same as those in the clock itself is challenging. This study numerically and analytically investigated a coupled system of a circadian clock and its output system, and then compared the fluctuations in the oscillation period of the two systems. We found that the amount of fluctuation in the output system is smaller than that in the circadian clock when the degradation rate of the molecules responsible for the output system was at typical values. The results obtained imply that the output system improves the accuracy of the circadian rhythm without the need for any special denoising processes.}

\section*{Author summary}
Animals, plants, and bacteria have approximately 24-hour cycles in their bodies, known as the circadian rhythm. The circadian clock, a molecular machinery generating daily rhythm, outputs a signal to peripherals that allows organisms to maintain their daily activities even when they are placed in a room that is isolated from time related information. The oscillation period of circadian rhythms are not constant, and varies slightly from day to day. However, the origin of fluctuation in circadian rhythm remains unknown because it is technically difficult to directly observe the dynamics of circadian clock and its signal transduction to output system in a single cell. Therefore, we theoretically examined the fluctuation in the system consisting of a central circadian clock and its output system. The numerical and analytical calculations showed that the circadian clock and its output system may exhibit different accuracies. Furthermore, the signal transduction from circadian clock to output system can reduce the fluctuations that circadian clocks originally possess. Thus, this study refers to a design principle that outputs the circadian rhythms of organisms in a more accurate manner.

\section*{Introduction}
Circadian rhythms are physiological phenomena that repeat approximately every 24 hours. The one prominent property of circadian rhythm is self-sustainability, that is, organisms can maintain their internal circadian oscillations under constant conditions \cite{Johnson:2004book}. Recent single-cell observations have revealed that the individual cells show self-sustained circadian oscillation even if they are uncoupled, such as a dissociation culture. Moreover, the single-cellular rhythms show fluctuations, that is, the period of circadian rhythm is irregular with a deviation \cite{Micklem.2021}. Such fluctuation in rhythms can be quantified using the coefficient of variation (CV), which is the standard deviation (SD) of the cycle-to-cycle periods divided by the average of those periods.
For example, the oscillation period of mammalian fibroblast cell populations is 24.38 hours on average and fluctuates with SD of 1.12 hours \cite{Li:2020}, thus, CV $\sim 0.05$. The single-cell measurements for the \textit{Arabidpsis} seedlings revealed CV $\sim 0.1$ \cite{Gould:2018}, which is close to mammalian cells. The circadian rhythm of prokaryote cyanobacteria is more accurate, with a period of 24.2 hours and SD of 0.12 hours, thus CV $\sim 0.005$ \cite{Mihalcescu:2004ch}.

The fluctuations in the circadian rhythms are partly because of fluctuations produced by the circadian molecular machinery \cite{Chabot:2007en}. The circadian clock system of any organism involves a transcription-translation feedback loop \cite{NobelPrize:2017}. Thus, the feedback loop in the circadian clock system contains a transcriptional noise, which causes the dispersion of circadian period \cite{Gonze:2002im, Nishino:2013ix}. In fact, the administration of a drug enhancing the transcriptional noise reduced the regularity of circadian oscillation \cite{Li:2020}.

The precision of circadian rhythms including the above mentioned examples have been observed using a reporter system with bioluminescence or fluorescence. For example, Li et al. \cite{Li:2020} used a reporter system wherein the clock-controlled Per2 promoter drove the expression of luciferase gene, resulting in daily rhythms of bioluminescence. Technically, fluctuations observed via these reporter systems are fluctuations of the output system. Certain theoretical studies state that the signal transduction can amplify fluctuations \cite{Shibata:2004, Shibata:2005th}. However, the association of fluctuations between the central circadian clock generating the circadian rhythms and the output system has been not addressed till date. This is because directly observing fluctuations of the central clock is difficult. Instead of experimental elucidation, we here theoretically addressed the question of whether the precision of circadian clock is improved or worsened by the downstream of the clock.

Several studies on the fluctuation in the oscillation periods have ever been performed theoretically. The CV in the period of synchronized coupled phase oscillators was analytically and numerically analyzed \cite{Kori:2012, Mori:2013ba}. Conversely, the coupling strength can be inferred based on the deviation of the period \cite{Mori:2022}. Moreover, Mori \& Mikhailov derived a formula describing period variability in general $N$-dimensional oscillatory systems. The authors revealed that the accuracy of rhythm depends on the choice of output variables, suggesting that the fluctuation in clock and output can be different \cite{Mori:2016dq}.

This study numerically examined a several mathematical models consisting of a noisy circadian clock and its output system. Moreover, we analytically calculated the fluctuations in oscillation period to understand how noise sources in the clock and output contribute to the variability in periods of the output rhythm.

\section*{Results}
\subsection*{Precision of the oscillation period in circadian clock with output system}
We examined the fluctuations of the oscillation period in the mathematical model for the circadian clock having its output system by employing the Goodwin model \cite{Goodwin.1965, KUROSAWA.2002} as a circadian clock. This model can represent a transcriptional-translational negative feedback loop of gene regulation underlying circadian clock\cite{NobelPrize:2017} and can show self-sustained oscillations. As a simple example of output system, we assumed a reporter protein such as luciferase or fluorescent protein, whose expression is driven by the circadian clock. The Goodwin oscillator with a reporter system is described as
\begin{equation}
\label{goodwin}
\begin{split}
    \frac{1}{T}\dot{u} &=\frac{1}{1+w^{m}}-k_{u}u+ \epsilon\sqrt{D_{u}}\xi_{u}(t), \\
    \frac{1}{T}\dot{v} &=u-k_{v}v, \\
    \frac{1}{T}\dot{w} &=v-k_{w}w, \\
    \dot{x} &=a+bw-k_xx + \epsilon\sqrt{D_{x}}\xi_{x}(t),
\end{split}
\end{equation}
where $u$ is the amount of mRNA, $v$ and $w$ are the amounts of clock protein in the cytoplasm and nucleus, respectively, $x$ represents the amount of reporter protein, coefficient $\epsilon \ll 1$ is a small parameter, $\sqrt{D_{u}}$ and $\sqrt{D_{x}}$ are the strength of noise of gene expression in the clock and output, respectively, and $\xi_{u}(t)$ and $\xi_{x}(t)$ are independent white Gaussian noise with $E[\xi (t)]= 0$ and $E[\xi(t)\xi(t')]=\delta(t-t')$, where $E$ represents the expectation. Further, $T$ is a parameter to scale the oscillation period to unity (day), $m$ is the Hill coefficient, $k_u$, $k_v$, and $k_w$ are the degradation rates of mRNA and clock proteins, $a$ is a basal production rate of $x$, $b$ is the coupling strength between clock and output, and $k_x$ is a degradation rate of a reporter protein.

We numerically solved Eq~\ref{goodwin} and then measured the variation of the oscillation periods of $w$ and $x$ (Fig 1A and Methods). Here, we defined the oscillation period as a peak-to-peak interval, which is a conventional measure in circadian studies. Consequently, based on the measured mean and SD of the oscillation periods, we obtained CV as an index of the accuracy of rhythmicity.
The numerical simulations revealed that the degradation rate of reporter protein, $k_x$, can modify the precision of the output rhythm (Fig 1B). Moreover, the fluctuation of the output system is less than that of the central circadian clock for a moderate value of $k_x$. The fluctuation in output period was minimized at $k_x \sim 10^{0.3}$.

\begin{figure}[!h]
\caption{{\bf Fluctuation in noisy Goodwin and phase model with output system.}
(A) Schematic for the Goodwin model and a clock-driven reporter system. The Goodwin model represents transcriptional and translational feedback loop generating circadian rhythms. The mRNA, $u$ and protein, $v$ can be expressed from a clock gene through the central dogma. A part of clock protein can move to the nucleus and then negatively regulate its own gene expression. The nucleic clock protein $w$ controls a promoter that fuses a reporter gene, causing daily bioluminescence or fluorescence rhythm. The transcriptional noise can provide fluctuations to $u$ and $x$. To measure the fluctuation of the oscillation period in circadian clock and output system, we detected peaks of $w$ and $x$, respectively. Thereafter, the coefficient variations (CV) of time intervals between the peaks was calculated. (B) Fluctuation of the oscillation periods in Goodwin oscillator and output systems at different degradation rates of $k_x$. $a=1.0$, $b=1.0$, $\epsilon^2D_{u}=8.75\times10^{-6}$, and $\epsilon^2D_{x}=1.0\times10^{-6}$. The error bars denote the standard error. (C) Schematic for the phase model and a clock-driven reporter system. The fluctuations are measured based on the peaks of $\sin \theta(t)$ and $x(t)$. (D) Fluctuation of the oscillation periods in the phase oscillator and output systems at different degradation rates of $k_x$. $\epsilon^2D_{\theta}=3.0\times10^{-2}$ and $\omega=2\pi$. The values of the other parameters are the same as Fig 1A.}
\label{fig1}
\end{figure}

\subsection*{Precision of the oscillation period in a phase oscillator with output system}
To check the generality of the enhanced precision in the output system, we adopted a phase oscillator for the model for circadian clock (Fig 1C), which can be described by
\begin{equation}
\label{model}
\begin{split}
    \dot{\theta} &=\omega + \epsilon\sqrt{D_{\theta}}\xi_{\theta}(t), \\
    \dot{x}&=a+b\sin \theta-k_xx + \epsilon\sqrt{D_{x}}\xi_{x}(t),
\end{split}
\end{equation}
where $\theta(t)$ is the phase of the circadian clock modulo $2\pi$, $\omega$ is the angular frequency of the circadian clock, $\sqrt{D_{\theta}}$ and $\sqrt{D_{x}}$ are the strengths of noise in the clock and output system, respectively, and $\xi_{\theta}(t)$ and $\xi_{x}(t)$ are independent white Gaussian noise applied for the circadian clock and output system, respectively.


The numerical simulation reproduced the qualitatively similar results to the Goodwin model (Fig 1D). The CV of the oscillation period in the output system depended on the degradation rate $k_x$ and can be smaller than that for circadian clock. The minimum value of CV was given at $k_x\sim 10^{0.8}$, which is of the same order as that of Goodwin model. The fact that the signal transduction can enhance the precision of rhythm regardless of choice of model suggests this property is universal.


Moreover, a study showed the actual half-life of proteins in a living human cell ranged between 45 min to 22.5 hours with mean 9.0 hours\cite{Eden:2011}. The half-life of 9.0 hours corresponds to 1.8 in the degradation rate. Taking into account that the enhancement of precision occurred when $k_x \sim 1$ to $10$ in our models, the abundance of the protein with a typical degradation rate can oscillate more precisely than circadian clock.

\subsection*{Analytical calculation for the fluctuation in oscillation periods confirmed the enhancement of precision}
To understand the mechanism of precision enhancement of circadian oscillation, we performed analytical calculations based on the theory developed by Mori and Mikhailov \cite{Mori:2016dq}. This study analytically derived the oscillation precision for the general $N$-dimensional model, which includes Eq~\ref{goodwin} and Eq~\ref{model}. The theory in \cite{Mori:2016dq} focuses on the fluctuations in oscillation periods defined by checkpoints (See Methods). The theory considers the deviation of oscillation phase, $\delta\Theta$ and amplitude, $\delta\boldsymbol{h}$ of the entire system during a single cycle (Fig 2). Thus, the CV of the oscillation period can be divided into three components:
\begin{equation}
    \label{origin_CV}
    \mathrm{CV} = \frac{\epsilon}{\tau}\sqrt{R_{\Theta\Theta}+R_{\boldsymbol{h}\boldsymbol{h}}+2R_{\Theta \boldsymbol{h}}} + \boldsymbol{O}(\epsilon^2),
\end{equation}
where $R_{\Theta\Theta}$ is collective-phase diffusion, $R_{\boldsymbol{h}\boldsymbol{h}}$ is the auto-correlation of amplitude deviation, and $R_{\Theta \boldsymbol{h}}$ is cross correlation between the collective-phase shift and amplitude deviation.

\begin{figure}[!h]
\caption{{\bf Phase and amplitude of the phase oscillator with a reporter.} The sample path of $\boldsymbol{x}(t) \equiv \left[\begin{array}{cc}\theta(t)\\ x(t)\end{array} \right]$ is shown in $(\theta, x)$ phase plane. $\boldsymbol{p}(t)$ is the limit cycle under noise-less conditions, where $D_{\theta}=0$ and $D_{x}=0$. The deviance during a single cycle, $\boldsymbol{x}\left(t_{\mathrm{cp}} + \tau\right)-\boldsymbol{x}(t_{\mathrm{cp}})$ are decomposed towards two directions $\epsilon\delta\Theta\dot{\boldsymbol{p}}\left(t_{\mathrm{cp}}\right)$ and $\epsilon\delta\boldsymbol{h}$ referred to as corrective phase shift and amplitude deviation of the entire system, respectively. For the considering system, Eq ~\ref{model}, the direction of $\delta\boldsymbol{h}$ is $\boldsymbol{P}(t_{\mathrm{cp}})\boldsymbol{\phi}_1=\left[\begin{array}{cc} 0 \\ 1 \end{array} \right]$ (\cite{Mori:2016dq} and Method).}
\label{fig2}
\end{figure}

As Eq~\ref{model} is a specific case of this general model and analytically tractable one, the CV values for both the clock and output systems can be obtained (details in Methods). We can decompose the CV of clock and output with respective noise source
into six components using Eq~\ref{origin_CV}:
\begin{equation}
\label{cv_all}
    \mathrm{CV} =  \epsilon\sqrt{D_{\theta}\left(R^{\left(\theta\right)}_{\Theta\Theta}+R^{\left(\theta\right)}_{\boldsymbol{h}\boldsymbol{h}}+2R^{\left(\theta\right)}_{\Theta \boldsymbol{h}}\right)+D_{x}\left(R^{\left(x\right)}_{\Theta\Theta}+R^{\left(x\right)}_{\boldsymbol{h}\boldsymbol{h}}+2R^{\left(x\right)}_{\Theta\boldsymbol{h}}\right)},
\end{equation}
where $R^{(\theta)}_{**}$ and $R^{(x)}_{**}$ denote the fluctuation in period caused by noise in clock and output system, respectively. It should be noted that the noise in clock and output contribute the CV in the form of a sum.

We first calculated the fluctuation in the period of circadian central clock, $\mathrm{CV}_{\mathrm{clock}}$. The components in Eq~\ref{cv_all} are expressed as
\begin{equation}
\label{R_clock}
    R^{\left(\theta\right)}_{\Theta\Theta} = \frac{\tau^2}{2\pi\omega},\;
    R^{\left(\theta\right)}_{\boldsymbol{h}\boldsymbol{h}},
    R^{\left(\theta\right)}_{\Theta \boldsymbol{h}},
    R^{\left(x\right)}_{\Theta\Theta},
     R^{\left(x\right)}_{\Theta \boldsymbol{h}},
    R^{\left(x\right)}_{\boldsymbol{h}\boldsymbol{h}}= 0.
\end{equation}

This result is understandable from Fig 2; $\delta\boldsymbol{h}$ is orthogonal to $\theta$, which implies no contribution of $\delta\boldsymbol{h}$. Thus, we get
\begin{equation}
    \label{cv_clock}
    \mathrm{CV}_{\mathrm{clock}} =  \epsilon\sqrt{\frac{D_{\theta}}{2\pi\omega}}.
\end{equation}
The fact that $\mathrm{CV}_\mathrm{clock}$ negatively depends on $\omega$ is reasonable because more frequent oscillator receives less noise during one cycle. Further, Eq~\ref{cv_clock} can be trivially derived without the theory by Mori \& Mikhailov. The SD of phase after a single oscillation is $\epsilon\sqrt{D_\theta\tau}$. By assuming (SD of oscillation period) $= \frac{\tau}{2\pi}$ (SD of phase)\cite{Kori:2012}, the CV of a period is $\frac{\epsilon\sqrt{D_\theta\tau}}{2\pi}$, which is equivalent to Eq~\ref{cv_clock}.

We next calculated the CV of the oscillation periods of the output system, $\mathrm{CV}_{\mathrm{output}}$. The components in Eq~\ref{cv_all} are expressed as
\begin{equation}
\label{R_output}
    \begin{split}
        R^{\left(\theta\right)}_{\Theta\Theta} &= \frac{\tau^2}{2\pi\omega},\\
        R^{\left(\theta\right)}_{\boldsymbol{h}\boldsymbol{h}} &= \frac{\tau^2(1-e^{-\kappa})}{2\pi\omega\kappa\left(\kappa^2+4\pi^2\right)}\cdot \left( \kappa^2+2\pi\kappa\tan\Phi_{\mathrm{cp}}+2\pi^2+2\pi^2\tan^2\Phi_{\mathrm{cp}} \right),\\
        R^{\left(\theta\right)}_{\Theta\boldsymbol{h}} &= -\frac{\tau^2(1-e^{-\kappa})}{2\pi\omega\kappa\left(\kappa^2+4\pi^2\right)}\cdot\left(\kappa^2+2\pi\kappa\tan\Phi_{\mathrm{cp}}\right),\\
        R^{\left(x\right)}_{\Theta\Theta} &= 0,\\
        R^{\left(x\right)}_{\boldsymbol{h}\boldsymbol{h}} &= \frac{\tau(1-e^{-\kappa})}{4\pi^2b^2\kappa}\left(\kappa^2+4\pi^2\right)\left(1+\tan^2\Phi_{\mathrm{cp}}\right),\\
        R^{\left(x\right)}_{\Theta\boldsymbol{h}} &= 0,
    \end{split}
\end{equation}
where $\kappa$ is the scaled degradation rate defined as $\kappa \equiv 2\pi k_x/\omega$. $\Phi_\mathrm{cp}$ is the checkpoint for observing periods. Thus, we obtain
\begin{equation}
    \label{cv_output}
    \begin{split}
        \mathrm{CV}_{\mathrm{output}}        = & \epsilon \Biggr\{ \frac{D_{\theta}}{2\pi\omega}\left[1+\frac{1-e^{\kappa}}{\kappa\left(\kappa^2+4\pi^2\right)}\left(-\kappa^2-2\pi\kappa\tan\Phi_{\mathrm{cp}}+2\pi^2+2\pi^2\tan^2\Phi_{\mathrm{cp}}\right)\right] \\
        &+ \frac{D_{x}}{2\pi\omega}\cdot \frac{\left(1-e^{-\kappa}\right)\omega^2}{4\pi^2b^2\kappa}\left(\kappa^2+4\pi^2\right)\left(1+\tan^2\Phi_{\mathrm{cp}}\right) \Biggl\}^{\frac{1}{2}}.
    \end{split}
\end{equation}
As in the numerical simulations, analytically calculated $\mathrm{CV}_\mathrm{output}$ depends on the degradation rate $\kappa$. Interestingly, the fluctuation originated from the noise in circadian clock, $R_{**}^{(\theta)}$, does not depend $a$ and $b$. $\omega$ determines the scale of 
$\mathrm{CV}_\mathrm{output}$.

We visually checked the consistency of the analytical and numerical CV under the fixed clock frequency, $\omega=2\pi$ (rad/day). To clearly understand the transmission of noise applied to the clock, we hereafter considered the noise-free condition for output system, that is, $D_x = 0$.
Both Eq~\ref{cv_clock} for $\mathrm{CV}_\mathrm{clock}$
and
Eq~\ref{cv_output} for $\mathrm{CV}_\mathrm{output}$
successfully reproduced the numerical results (Fig 3A). The $\mathrm{CV}_\mathrm{output}$ has a minimum value at $k_x \sim 10$ day${}^{-1}$, that is, $\kappa= 2 \pi \frac{ k_x}{\omega}\sim 10$. This fact implies that the precision of the circadian rhythms can be enhanced in the downstream if the values of the frequency and decay rate are at the same order. When $k_x \rightarrow \infty$, the value of $\mathrm{CV}_\mathrm{output}$ approaches the value of $\mathrm{CV}_\mathrm{clock}$, meaning that noise in the clock completely transmits to the downstream under the condition of frequent turnover of output protein. When $k_x \rightarrow 0$, $\mathrm{CV}_\mathrm{output}$ converges to $\epsilon\sqrt{\frac{D_{\theta}}{4\pi\omega}\left(3+\tan^2\Phi_\mathrm{cp}\right)}$.
Such behaviors hold qualitatively regardless of the choice of checkpoint $\Phi_\mathrm{cp}$ (right panel in Fig 3A) although $\mathrm{CV}_{\mathrm{output}}$ depends on $\Phi_\mathrm{cp}$ as in Eq~\ref{cv_output}.

We also visually checked the $\omega$-dependence of CV (Fig 3B). As noted above, $\omega$ controls the absolute values of fluctuation rather than function form of CV. For the larger value of $\omega$, the circadian clock and its output showed smaller fluctuations. This result is consistent with the observation that cultured mammalian cells with longer circadian period show less precise rhythm \cite{Li:2020,Nikhil.2020}.

To elucidate the enhancement of oscillation precision in the downstream of the clock, we visually decomposed $\mathrm{CV}_\mathrm{output}$ into $R^{\left(\theta\right)}_{\Theta\Theta}$, $R^{\left(\theta\right)}_{\boldsymbol{h}\boldsymbol{h}}$, and $R^{\left(\theta\right)}_{\Theta\boldsymbol{h}}$ where $\omega=2\pi$ and $\Phi_{\mathrm{cp}}=0$ (Fig 3C). $R^{\left(\theta\right)}_{\Theta\Theta}$ is positive and independent of $k_x$ because it originates from the fluctuation of the central clock.  $R^{(\theta)}_{\boldsymbol{h}\boldsymbol{h}}$ is also a positive and monotonically decreasing function of $k_x$.
On the contrary, $R^{\left(\theta\right)}_{\Theta\boldsymbol{h}}$ is negative for any $k_x$ and a bowl-shaped function that has the minimum value at approximately $k_x=10$ when $\Phi_{\mathrm{cp}} = 0$.
We confirmed that $R^{\left(\theta\right)}_{\Theta\boldsymbol{h}}$ is negative for a broad range of the value of $\Phi_{\mathrm{cp}}$ at the typical value of the protein degradation rate in human, $k_x=1.8$.
Thus, negative correlation of phase shift and amplitude variation can reduce output fluctuation in actual biological clocks.

Moreover, it appears that the fluctuation numerically measured by the peak-to-peak cycle is coincident with $R^{\left(x\right)}_{\Theta\Theta}+2R^{\left(x\right)}_{\Theta\boldsymbol{h}}$ (Fig 3D). This agreement suggests that peak-to-peak period can be a more precise measure than period measured based on a checkpoint because of independence of amplitude deviation, $R^{\left(x\right)}_{\boldsymbol{hh}}$.

\begin{figure}[!h]
\caption{{\bf Analytical calculation for the fluctuation of the circadian clock and output system.}
(A) Comparison of numerical and analytical calculations for the precision of cycles. We measured the oscillation period based on two different checkpoints, that is, we defined the two thresholds at half ($\Phi_\mathrm{cp}=0$, left panel) and three-fourths ($\Phi_\mathrm{cp}=\frac{\pi}{6}$, right panel) of the oscillatory range of $\sin \theta(t)$ or $x(t)$. We detected the times at which $\sin \theta(t)$ or $x(t)$ passed the thresholds from below to above. The period is defined as the interval of the detected times. The parameters in Fig 3 are set to $\omega = 2\pi$ and $\epsilon^2D_{\theta}=3.0\times10^{-2}$. (B) Dependency of the precision of cycles on $\omega$. (C) The three component of CV, $R^{\left(\theta\right)}_{\Theta\Theta}, R^{\left(\theta\right)}_{\boldsymbol{h}\boldsymbol{h}}$, and $R^{\left(\theta\right)}_{\Theta\boldsymbol{h}}$. (D) Numerically-calculated fluctuations of peak-to-peak cycles and analytical calculations, $\frac{\epsilon}{\tau}\sqrt{D_{\theta}\left(R^{\left(\theta\right)}_{\Theta\Theta}+2R^{\left(\theta\right)}_{\Theta\boldsymbol{h}}\right)}$ for the precision of peak-to-peak cycles in output system. $\Phi_\mathrm{cp}=0$.}
\label{fig3}
\end{figure}

\section*{Discussion}


We numerically showed that the circadian clock consisting of transcriptional-translational feedback loop and its downstream output differ in the amount of fluctuations of the oscillation period. Moreover, the degradation rate of the molecules in the output system controlled the amount of fluctuation. In a range of the degradation rate, fluctuations in output system was smaller than those of the circadian clock generating the rhythm. Furthermore, this reduction in fluctuation was also reproduced by a phase model and was consistent with the analytical results. The analytical calculations decomposed the amount of fluctuation into six components based on the noise sources and the directions on the phase plane the whole system. One of the components, the negative cross-correlation between the phase and amplitude of the fluctuations originated from the noise applied to the central clock, $R^{(\theta)}_{\Theta\boldsymbol{h}}$, contributed to the reduction of fluctuation in output. When the component cancel out the other terms, the output system became more precise than the central circadian clock (Fig 4).

\begin{figure}[!h]
\caption{{\bf Fluctuations in circadian system driven by noise}}
Schematic for contributions of noise in circadian clock and output system to fluctuations.
Noise applied to circadian clock, $D_{\theta}\xi_{\theta}$, provides the fluctuation in the circadian clock, $D_{\theta}R^{(\theta)}_{\Theta\Theta}$.
This fluctuation is transmitted to output system and further yields the fluctuations, $D_{\theta}R^{(\theta)}_{\Theta\Theta}$, $2D_{\theta}R^{(\theta)}_{\Theta\boldsymbol{h}}$ and $D_{\theta}R^{(\theta)}_{\boldsymbol{h}\boldsymbol{h}}$. In addition, noise applied to the output system, $D_{x}\xi_{x}$ directly provides the fluctuation in output system, $D_{x}R^{(x)}_{\boldsymbol{h}\boldsymbol{h}}$.
Thus, the fluctuation in the output system consists of the four contributions while that in the circadian clock has only $D_{\theta}\xi_{\theta}$. However, the output system can show smaller fluctuation than the circadian clock because $2D_{\theta}R^{(\theta)}_{\Theta\boldsymbol{h}}$ originating from the circadian clock, is negative under proper values of $k_x$ and $\Phi_{\mathrm{cp}}$.Thus, the transmission of time information can reduces the fluctuations in the downstream system.
\label{fig4}
\end{figure}

These results indicate that the output system transmits the time information to the physiological level and embeds the function of controlling the amount of fluctuation. This provides a new perspective of the output system in circadian molecular machinery. The circadian clock controls numerous physiological functions \cite{Johnson:2004}. Our theoretical finding implicitly suggests that different physiological rhythms have different precision of rhythms. The rate constants in output system can be optimized if the output requires precise time regulations.

Conversely, the circadian machinery in a cell can be noisier than the output system. The dynamics of the clock proteins at a single cell level has never been observed without reporter system. However, if this is possible, the heart of circadian machinery would be more fragile than expected. For example, KaiC phosphorylation in cyanobacteria is considered as a central circadian oscillator \cite{Nakajima:2005jc}. If the level of KaiC phosphorylation in a cell could be measured non-invasively, such as fluorescence correlation spectroscopy \cite{Goda:2012}, the intact clock dynamics may exhibit noisy behaviour.


Among the parameters in the output system, only $k_x$ was involved in the fluctuations transmitted from noisy circadian clock. This theoretical finding suggests that protein degradation controls the precision of the rhythm rather than protein synthesis. Moreover, we showed the optimized degradation rate $k_x$ exists in a range of 1 to 10, which corresponds to 1.7 to 17 hours in half-life. This range overlaps the actual half-life of proteins in a living human cell \cite{Eden:2011}, suggesting that certain proteins can contribute to the enhancement of precision in human circadian rhythm.

We considered a model consisting of two factors: the clock and the output system. A three-factor model, in which an additional factor is connected to the downstream of this model, may result in smaller fluctuations. This is because the signal transduction, which is the key to reducing fluctuations, occurs twice. In real circadian systems, there are many factors connecting the clock and the physiological output\cite{Dibner:2010,Bell:2005}. Such a multistep and complex biological network may have been optimizely  constructed through evolution to obtain more accurate rhythms.

This study was performed from a theoretical perspective. Our conclusion that the output system can contribute to the precision of rhythmic period needs experimental validations. In fact, we can experimentally control the value of degradation rate, $k_x$, by the addition of degradation tag to the reporter protein \cite{Andersen:1998} or introduction of an inducible protein degradation system \cite{Daniel:2018}. In addition, dilution due to cell division can also be an alternative parameter of $k_x$ because it is involved in the turnover of a fluorescent protein in a cell and can effectively control the degradation rate. Furthermore, nutritional or thermal conditions also might control the precision of circadian rhythms through alteration of cell division rate.

Moreover, although we have treated Eq~\ref{model} as the model describing circadian rhythms till now, this abstract model can be regarded as a representative for other biological rhythm, such as rhythmic firing of neurons and compression of cardiomyocytes. These oscillations would require more precision for proper biological functions. Therefore, the mechanism of output systems that make rhythms more precise could be discovered in another biological oscillators other than circadian rhythms. In addition, the concept of output-enhanced precision can be applied for synthetic oscillatory genetic circuits \cite{Purcell:2010}. A more accurate synthesized biological clock can be realized by optimizing output system. Thus, this study provides a design principle for accurate oscillating biological circuits.

\section*{Acknowledgments}
We thank I. Mihalcescu (Universtite Grenoble Alpes) for a fruitful discussion. This work was supported in part by Japan Society for the Promotion of Science KAKENHI Grants JP18H05474 (H.I.), JP19K03663 (F. M.) and JP22K03453 (F. M.).

\section*{Author Contributions}
\noindent\textbf{Conceptualization:} Hiroshi Ito, Fumito Mori.\\
\noindent\textbf{Analytic Calculation:} Hotaka Kaji, Fumito Mori, Hiroshi Ito.\\
\noindent\textbf{Numerical simulation:} Hotaka Kaji.\\
\noindent\textbf{Writing – original draft:} Hotaka Kaji, Fumito Mori, Hiroshi Ito.\\

\bibliography{outputnoise.bib}
\section*{Methods}

\subsection*{Analytical calculation for the precision of circadian rhythms}
Based on the theory developed by Mori \& Mikhailov \cite{Mori:2016dq}, we analytically derived the fluctuation in the oscillation period of the model Eq~\ref{model}. The authors considered a general $N$-dimensional system in the presence of noise expressed as
\begin{equation*}
    \frac{d\boldsymbol{x}}{dt}=
    \boldsymbol{f}[\boldsymbol{x}(t)]+
    \epsilon \boldsymbol{G}[\boldsymbol{x}(t)]\boldsymbol{\xi}(t),
\end{equation*} where $\boldsymbol x(t)$ is a vector with $N$ elements, $\boldsymbol{f}$ represents the $N$-dimensional oscillatory system that generates a limit-cycle solution
with period $\tau=2\pi/\omega$
under noise-less conditions, coefficient $\epsilon \ll 1$ is a small parameter, $\boldsymbol{G}[\boldsymbol{x}(t)]$ is a $N \times N$ diagonal matrix, and $\boldsymbol{\xi}(t)$ represents a vector of additive noise which holds $E[\xi_i ] = 0$ and $E[\xi_i (t_1)\xi_j (t_2)] = \delta_{ij}\delta(t_1 - t_2)$.

where $\boldsymbol{f}[\boldsymbol{x}(t)]=
    \left[\begin{array}{cc}
         f_{\theta}(\theta, x)\\
         f_{x}(\theta, x)
    \end{array}\right]=
    \left[\begin{array}{cc}
         \omega\\
        a + b\sin\theta-k_xx
    \end{array}\right]$, $\boldsymbol{G}[\boldsymbol{x}(t)]=\left[\begin{array}{cc}
    \sqrt{D_\theta} & 0 \\
    0 & \sqrt{D_x}
\end{array}
\right]$, and $\boldsymbol{\xi}(t)=\left[\begin{array}{cc}
     \xi_\theta(t)  \\
     \xi_x(t)
\end{array}\right]$.
The solution for Eq~\ref{model} without noise ($\epsilon=0$) is
\begin{equation*}
\begin{split}
    \theta(t)&=\omega t+\theta(0)  \mod 2\pi,\\
    x(t)&=\frac{a}{k_x}+\frac{b}{k_x^2+\omega^2}\left[k_x\sin(\omega t+\theta(0))-\omega\cos(\omega t+\theta(0))\right]\\
    &+\left[x(0)-\frac{a}{k_x}-\frac{b}{k_x^2+\omega^2}(k_x\sin \theta(0)-\omega\cos \theta(0))\right]e^{-k_xt}.
\end{split}
\end{equation*}
Thus, if the initial condition is set to $(\theta(0), x(0))=\left(0, \frac{a}{k_x}-\frac{b\omega}{k_x^2+\omega^2}\right)$, a limit cycle solution $\boldsymbol{p}(t)$ is obtained as
\begin{equation}
  \label{p}
  \boldsymbol{p}(t)
    =\left[
    \begin{array}{cc}
    \omega t \mod 2\pi \\
    \frac{a\tau}{\kappa}+\frac{b\tau}{\sqrt{\kappa^2+4\pi^2}}\sin\left(\omega t-\tan^{-1}\left(\frac{2\pi}{\kappa}\right)\right)
    \end{array}
    \right],
\end{equation}
where $\kappa=2\pi k_x/\omega$. Note that $\boldsymbol{p}(t)$ satisfies $\boldsymbol{p}(t)=\boldsymbol{p}(t+\tau)$, where the oscillation period without noise is $\tau = 2\pi/\omega$.

Suppose that $\boldsymbol{x}(t)$ is located near the limit cycle, that is, $\boldsymbol{x}(t)$ can be written as
\begin{equation*}
  \boldsymbol{x}(t) = \boldsymbol{p}(t) + \epsilon\boldsymbol{z}(t) + \boldsymbol{O}(\epsilon^2),
\end{equation*}
where $\left|\left|\boldsymbol{z}\right|\right| \ll \epsilon^{-1}$. Then, $\boldsymbol{z}(t)$ obeys the linearized equation
\begin{equation*}
  \frac{d\boldsymbol{z}}{dt} = \boldsymbol{\Gamma}(t)\boldsymbol{z}(t) + \boldsymbol{G}[\boldsymbol{p}(t)]\boldsymbol{\xi}(t),
\end{equation*}
where the Jacobian matrix $\boldsymbol\Gamma(t)$ is expressed as
\begin{equation*}
    \boldsymbol\Gamma(t)
     =\left[
    \begin{array}{cc}
    \frac{\partial f_{\theta}}{\partial \theta} & \frac{\partial f_{\theta}}{\partial x} \\
    \frac{\partial f_{x}}{\partial \theta} & \frac{\partial f_{x}}{\partial x} \\
    \end{array}
    \right]_{\boldsymbol{x}(t)=\boldsymbol{p}(t)}
    =\left[
    \begin{array}{cc}
    0 & 0 \\
    b\cos\omega t & -\frac{\kappa}{\tau} \\
    \end{array}
    \right].
\end{equation*}

\noindent
Then, we considered the unperturbed system
\begin{equation}
    \label{z1}
    \frac{d\boldsymbol{z}}{dt} = \boldsymbol{\Gamma}(t)\boldsymbol{z}(t)=
    \left[
    \begin{array}{cc}
    0 & 0 \\
    b\cos\omega t & -\frac{\kappa}{\tau} \\
    \end{array}
    \right]\left[
    \begin{array}{cc}
    z_1(t) \\
    z_2(t)\\
    \end{array}
    \right].
\end{equation}
Eq~\ref{z1} can be solved with the initial value of $\boldsymbol{z}(0)$ as

\begin{equation*}
  \begin{aligned}
    \label{z}
    \boldsymbol z(t)=
    \left[
    \begin{array}{cc}
    1 & 0\\
    \frac{b\tau}{\kappa^2+4\pi^2}\left\{\sqrt{\kappa^2+4\pi^2}\cos\left(\omega t-\tan^{-1}\left(\frac{2\pi}{\kappa}\right)\right)-\kappa e^{-\frac{\kappa}{\tau}t}\right\} & e^{-\frac{\kappa}{\tau}t}
    \end{array}
    \right]\left[
    \begin{array}{cc}
    z_1(0) \\
    z_2(0)\\
    \end{array}
    \right].\\
  \end{aligned}
\end{equation*}
Moreover, we can rewrite $\boldsymbol{z}(t)$ as
\begin{equation*}
    \boldsymbol{z}(t) = \boldsymbol{U}(t)\boldsymbol{U}(0)^{-1}\boldsymbol{z}(0),
\end{equation*}
where $\boldsymbol{U}(t)$ is called a fundamental matrix solution, defined as
\begin{equation*}
    \label{U}
    \boldsymbol U(t)=\left[
    \begin{array}{cc}
    1 & 0\\
    \frac{b\tau}{\kappa^2+4\pi^2}\left\{\sqrt{\kappa^2+4\pi^2}\cos\left(\omega t-\tan^{-1}\left(\frac{2\pi}{\kappa}\right)\right)-\kappa e^{-\frac{\kappa}{\tau}t}\right\} & e^{-\frac{\kappa}{\tau}t}
    \end{array}
    \right].
\end{equation*}
 Here, the constant matrix $\boldsymbol{B}$ and a periodic matrix function $\boldsymbol{P}(t)$ are introduced. They are defined as
\begin{equation*}
    \begin{split}
        \exp(\tau\boldsymbol{B}) &\equiv \boldsymbol{U}(0)^{-1}\boldsymbol{U}(\tau),\\
        \boldsymbol{P}(t)&\equiv \boldsymbol{U}(t) e^{-t\boldsymbol{B}}.
    \end{split}
\end{equation*}
For our model,  $\boldsymbol{B}$ and $\boldsymbol{P}(t)$ are
\begin{equation*}
  \boldsymbol B=
  \left[
  \begin{array}{cc}
  0 & 0\\
  \frac{b\kappa^2}{\kappa^2+4\pi^2} & -\frac{\kappa}{\tau}
  \end{array}
  \right],
\end{equation*}
\begin{equation*}
\quad\boldsymbol P(t)=
\left[
\begin{array}{cc}
1 & 0\\
\frac{b\tau}{\kappa^2 + 4\pi^2}\left\{\sqrt{\kappa^2+4\pi^2}\cos\left(\omega t-\tan^{-1}\left(\frac{2\pi}{\kappa}\right)\right)-\kappa\right\} & 1
\end{array}
\right].
\end{equation*}
Thus, we obtain the right and left eigenvector of $\boldsymbol B$, $\boldsymbol{\phi}$ and ${}^t\boldsymbol{\psi}$, respectively, where the superscript $t$ implies the transposition.
According to the Floquet theory \cite{coddington1326linear}, the eigenvalues of $\boldsymbol B$ are referred to as Floquet exponents, and one of the Floquet components should be zero because we consider the model showing the limit cycle oscillations. In fact, if the eigenvalues of $\boldsymbol B$ are $\lambda_0=0$, and $\lambda_1=-\kappa/\tau$.
For $\lambda_0=0$, we obtain
\begin{equation*}
    \boldsymbol{\phi}_0 =\left[
    \begin{array}{cc}
    \omega \\
    \frac{2\pi b\kappa}{\kappa^2+4\pi^2}
    \end{array}
    \right],\quad
    {\boldsymbol\psi}_0 =\left[
    \begin{array}{cc}
    \frac{1}{\omega} \\
    0
    \end{array}
    \right].
\end{equation*}
For $\lambda_1=-\kappa/\tau$, we obtain
\begin{equation*}
    \boldsymbol{\phi}_1 =\left[
    \begin{array}{cc}
    0 \\
    1
    \end{array}
    \right], \quad
    {\boldsymbol\psi}_1 =\left[
    \begin{array}{cc}
    \frac{-b\tau\kappa}{\kappa^2+4\pi^2} \\
    1
    \end{array}
    \right].
\end{equation*}

According to \cite{Mori:2016dq}, the CV of the oscillation periods can be calculated based on the three components,
\begin{equation}
\label{originalcv2}
    \mathrm{CV} = \frac{\epsilon}{\tau}\sqrt{R_{\Theta\Theta}+R_{\boldsymbol{h}\boldsymbol{h}}+2R_{\Theta\boldsymbol{h}}} + \boldsymbol{O}(\epsilon^2),
\end{equation}
where $R_{\theta\theta}$,$R_{\boldsymbol{hh}}$, and $R_{\theta\boldsymbol{h}}$ correspond to collective-phase diffusion, the auto-correlation of amplitude deviation, and cross correlation between the collective-phase shift and the amplitude deviation, respectively. These components can be expressed using $\boldsymbol{P}(t)$, $\boldsymbol{\phi}_{0,1}$, and $\boldsymbol{\psi}_{0,1}$, as following,

\begin{equation}
\label{cv_general}
\begin{split}
    R_{\Theta\Theta}&=\int_0^\tau\!{}^t\!{\boldsymbol\psi_0}\boldsymbol P(t)^{-1}\boldsymbol G[\boldsymbol p(t)]^2\hspace{2pt} {}^t\!\boldsymbol P(t)^{-1}\boldsymbol\psi_0 dt,\\
    R_{\boldsymbol{h}\boldsymbol{h}}&=\sum\limits_{j=1}^{N-1}\sum\limits_{k=1}^{N-1}[2-\exp(\lambda_j\tau)-\exp(\lambda_k\tau)]\frac{[\boldsymbol P(t_\mathrm{cp}){\boldsymbol\phi_j}]_l[\boldsymbol P(t_\mathrm{cp}){\boldsymbol\phi_k}]_l}{[\dot{\boldsymbol p}(t_{\mathrm{cp}})]^2_l} \exp[(\lambda_j+\lambda_k)t_\mathrm{cp}]\\
    & \times\bigg\{ \int_0^{t_\mathrm{cp}}\!\exp[-(\lambda_j+\lambda_k)t]\hspace{2pt}{}^t\!{\boldsymbol\psi_j}\boldsymbol P(t)^{-1}\boldsymbol G[\boldsymbol p(t)]^2\hspace{2pt}{}^t\!\boldsymbol P(t)^{-1}{\boldsymbol\psi_k} dt \\
    &+\frac{\exp[(\lambda_j+\lambda_k)\tau]}{1-\exp[(\lambda_j+\lambda_k)\tau]}\int_0^\tau\!\exp[-(\lambda_j+\lambda_k)t]\hspace{2pt}{}^t\!{\boldsymbol\psi_j}\boldsymbol P(t)^{-1}\boldsymbol G[\boldsymbol p(t)]^2\hspace{2pt}{}^t\!\boldsymbol P(t)^{-1}{\boldsymbol\psi_k}dt \bigg\}, \\
    R_{\Theta\boldsymbol{h}}&=\sum\limits_{j=1}^{N-1}\exp(\lambda_j\tau)\frac{[\boldsymbol P(t_\mathrm{cp}){\boldsymbol\phi_j}]_l}{[\dot{\boldsymbol p}(t_\mathrm{cp})]_l}\\
    &\times\int_0^\tau\!\exp(-\lambda_j t)\hspace{2pt}{}^t\!{\boldsymbol\psi_0}\boldsymbol P(t_\mathrm{cp}+t)^{-1}\boldsymbol G[\boldsymbol p(t+t_\mathrm{cp})]^2\hspace{2pt}{}^t\!\boldsymbol P(t_\mathrm{cp}+t)^{-1}{\boldsymbol\psi_j}dt,
  \end{split}
\end{equation}
where $N = 2$ is the dimension of the considered systems, $[\boldsymbol x]_l$ denotes the $l$th element of the vector $\boldsymbol x$, $t_\mathrm{cp}$ is the time determined based on choice of checkpoint, which is defined at the subsection of "Measurement of fluctuation in peak-to-peak periods".

To understand the extent to which each noise source contributes to CV, we decompose $\boldsymbol{G}[\boldsymbol{p}(t)]^2$ as follows,
\begin{equation}
\label{G}
\boldsymbol{G}[\boldsymbol{p}(t)]^2=D_\theta\left[\begin{array}{cc}
    1 & 0 \\
    0 & 0
\end{array}
\right] + D_x\left[\begin{array}{cc}
    0 & 0 \\
    0 & 1
\end{array}
\right].
\end{equation}
The first and second terms of RHS of Eq~\ref{G} represent the noise strength applied to $\theta$ and $x$, respectively. Considering the linearity of Eq~\ref{cv_general}, we can rewrite Eq~\ref{originalcv2} as
\begin{equation}
    \mathrm{CV} = \frac{\epsilon}{\tau}\sqrt{
    D_{\theta}R^{\left(\theta\right)}_{\Theta\Theta} + D_{x}R^{\left(x\right)}_{\Theta\Theta}+
    D_{\theta}R^{\left(\theta\right)}_{\boldsymbol{h}\boldsymbol{h}} + D_{x}R^{\left(x\right)}_{\boldsymbol{h}\boldsymbol{h}} +
    2D_{\theta}R^{\left(\theta\right)}_{\Theta\boldsymbol{h}} + 2D_{x}R^{\left(x\right)}_{\Theta\boldsymbol{h}}},
\end{equation}
where the values of $R^{\left(\theta\right)}_{**}$ and $R^{\left(x\right)}_{**}$ are calculated under $\boldsymbol{G}[\boldsymbol{p}(t)]^2=
\left[
\begin{array}{cc}
    1 & 0 \\
    0 & 0
\end{array}
\right]$ and
$\left[
\begin{array}{cc}
    0 & 0 \\
    0 & 1
\end{array}
\right]$, respectively, which is equivalent to Eq~\ref{cv_all}. Using Eq~\ref{cv_general} and setting $l=1$ or $l=2$ for the CV of the circadian clock or the output system, we obtain Eq~\ref{R_clock} and Eq~\ref{R_output}.

\subsection*{Numerical Simulation}
Equation(\ref{goodwin}) and (\ref{model}) were numerically solved for 1000 cycles with Euler's method where $\Delta t=1.0\times10^{-4}$. For Eq(\ref{goodwin}), the initial value of ($u(0), \; v(0), \; w(0), \; x(0)$) was set to ($ 0, \; 0, \; 0, \; 0$). Whereas, for Eq (\ref{model}), the initial value of ($\theta(0), \; x(0)$) was $\left(0, \; \frac{1}{k_x}-\frac{2\pi}{k_x^2+4\pi^2}\right)$, which is on the limit cycle, Eq~\ref{p}. We numerically obtained the CV in the simulation, and repeated it 100 times to obtain the average of the CV values in Fig 1, and 10 times in Fig 3.

\subsection*{Measurement of fluctuation in peak-to-peak periods}
For Fig 1B, we measured the fluctuations in the peak-to-peak periods of the clock and the output. We detected the peaks of $\sin\theta(t)$ and $x(t)$ using the function in the module of Python3, scipy.signal.argrelmax.

\subsection*{Measurement of fluctuation in periods determined by a checkpoint}
For Fig 3A, we set two different checkpoints to measure the oscillation periods of $\sin\theta(t)$ and $x(t)$. The limit cycle for our model under noise-free conditions, Eq~\ref{p}, can be rewritten as
\begin{equation*}
    \begin{split}
        \sin\theta(t) &=\sin\omega t,\\
        x(t)&=\frac{a\tau}{\kappa} + \frac{b\tau}{\sqrt{\kappa^2+4\pi^2}}\sin\left(\omega t - \tan^{-1}\left(\frac{2\pi}{\kappa}\right) \right),\\
    \end{split}
\end{equation*}
or,
\begin{equation*}
    \begin{split}
        \sin\theta(t) &=\sin\Phi^\mathrm{clock}(t),\\
        x(t)&=\frac{a\tau}{\kappa} + \frac{b\tau}{\sqrt{\kappa^2+4\pi^2}}\sin\Phi^\mathrm{output}(t),\\
    \end{split}
\end{equation*}
by introducing the apparent phase on the limit cycle, $\Phi^\mathrm{clock}(t)\equiv\omega t$ and $\Phi^\mathrm{output}(t)\equiv \omega t - \tan^{-1}\left(2\pi/\kappa\right)$. Note that $\Phi^\mathrm{output}(t)$ is dependent on the values of $\omega$ and $\kappa$, that is, $k_x (=\kappa\omega/2\pi)$.

Then, we employed the two checkpoints: $\Phi_\mathrm{cp}= 0\pi$ and
$\frac{\pi}{6}$.
In other words,
we provided the two thresholds, which are at the half and three-fourths value points of the oscillatory range of $\sin \Phi^\mathrm{clock} (t)$ and $\sin \Phi^\mathrm{output}(t)$, respectively.

To perform the numerical analysis, we recorded the times at which $\sin \theta(t)$ and $x(t)$ passed the thresholds, $\sin\Phi_\mathrm{cp}$
and $\frac{a\tau}{\kappa}+\frac{b\tau}{\sqrt{\kappa^2+4\pi^2}}\sin\Phi_\mathrm{cp}$, from below to above, respectively. Subsequently, we calculated the value of CV based on the time intervals. For large noise cases, $x(t)$ may not exceed these thresholds.
In our simulations,
when $x(t)$ did not exceed the threshold during each cycle even once,
the entire time series was discarded. The Python code for this procedure is provided through the website in Github below: \url{https://github.com/hito1979/outputnoise}.

For analytical calculation of Eq~\ref{R_clock} and Eq~\ref{R_output},
we used Eq~\ref{cv_general}, which contains the parameter
 $t_\mathrm{cp}$.
  $t^\mathrm{clock}_\mathrm{cp}$
 and
 $t^\mathrm{output}_\mathrm{cp}$
 were determined using
 common $\Phi_\mathrm{cp}$:
 \begin{equation*}
    \begin{split} t^\mathrm{clock}_\mathrm{cp} &= \frac{\Phi_\mathrm{cp}}{\omega},\\
        t^\mathrm{output}_\mathrm{cp} &= \frac{\tan^{-1}(\frac{2\pi}{\kappa})+\Phi_\mathrm{cp}}{\omega}.
    \end{split}
\end{equation*}

\includepdf[pages=-]{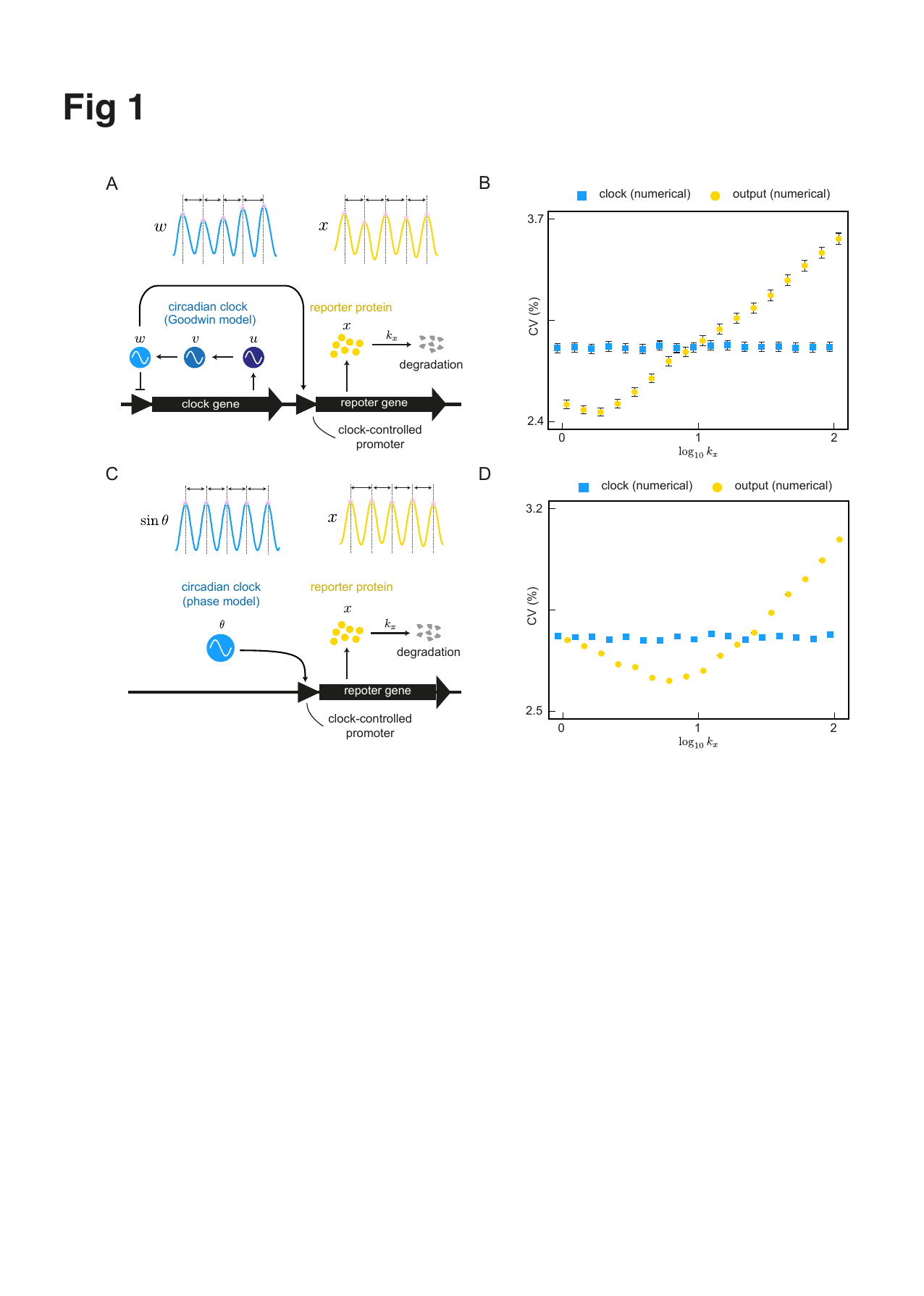}
\includepdf[pages=-]{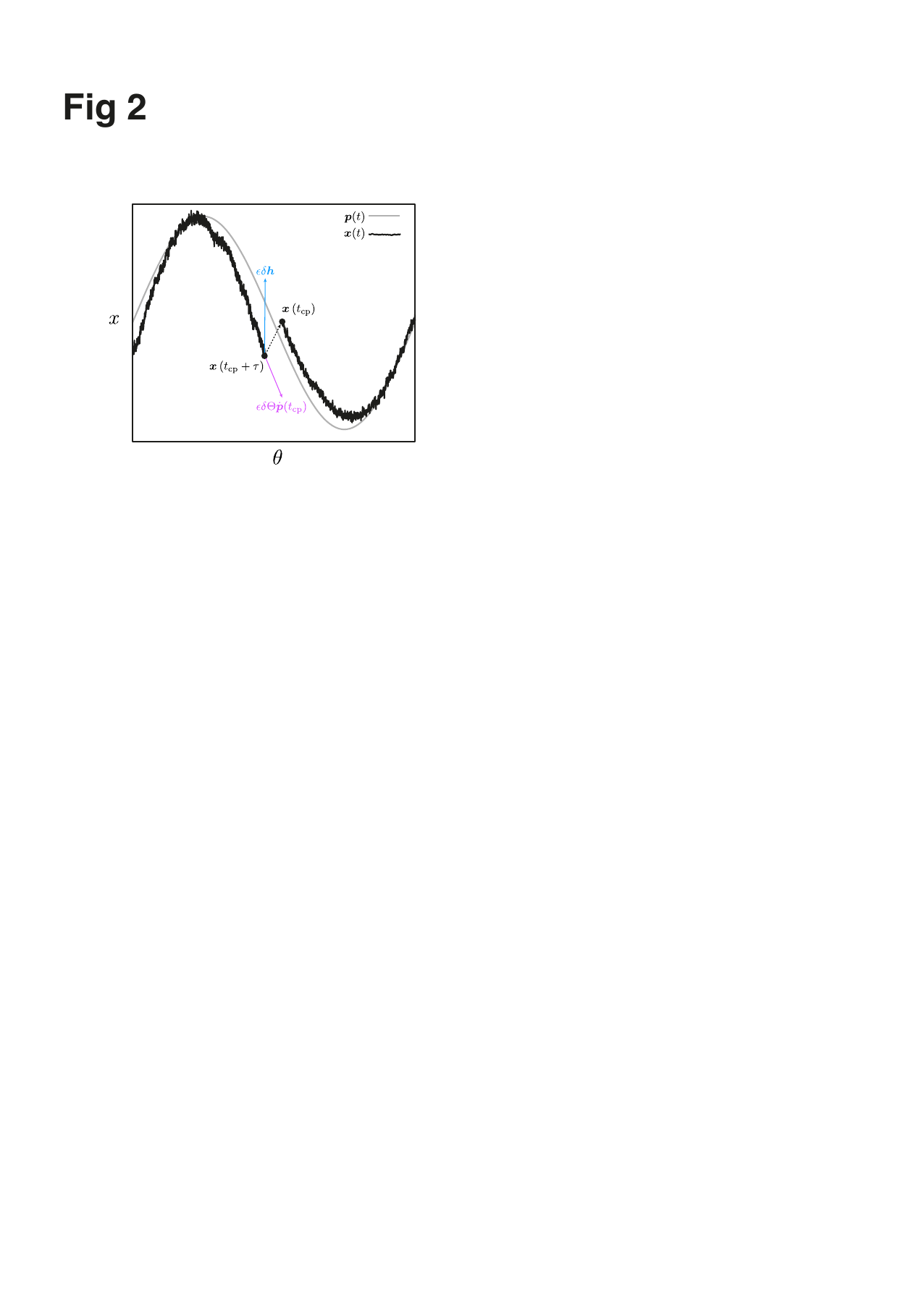}
\includepdf[pages=-]{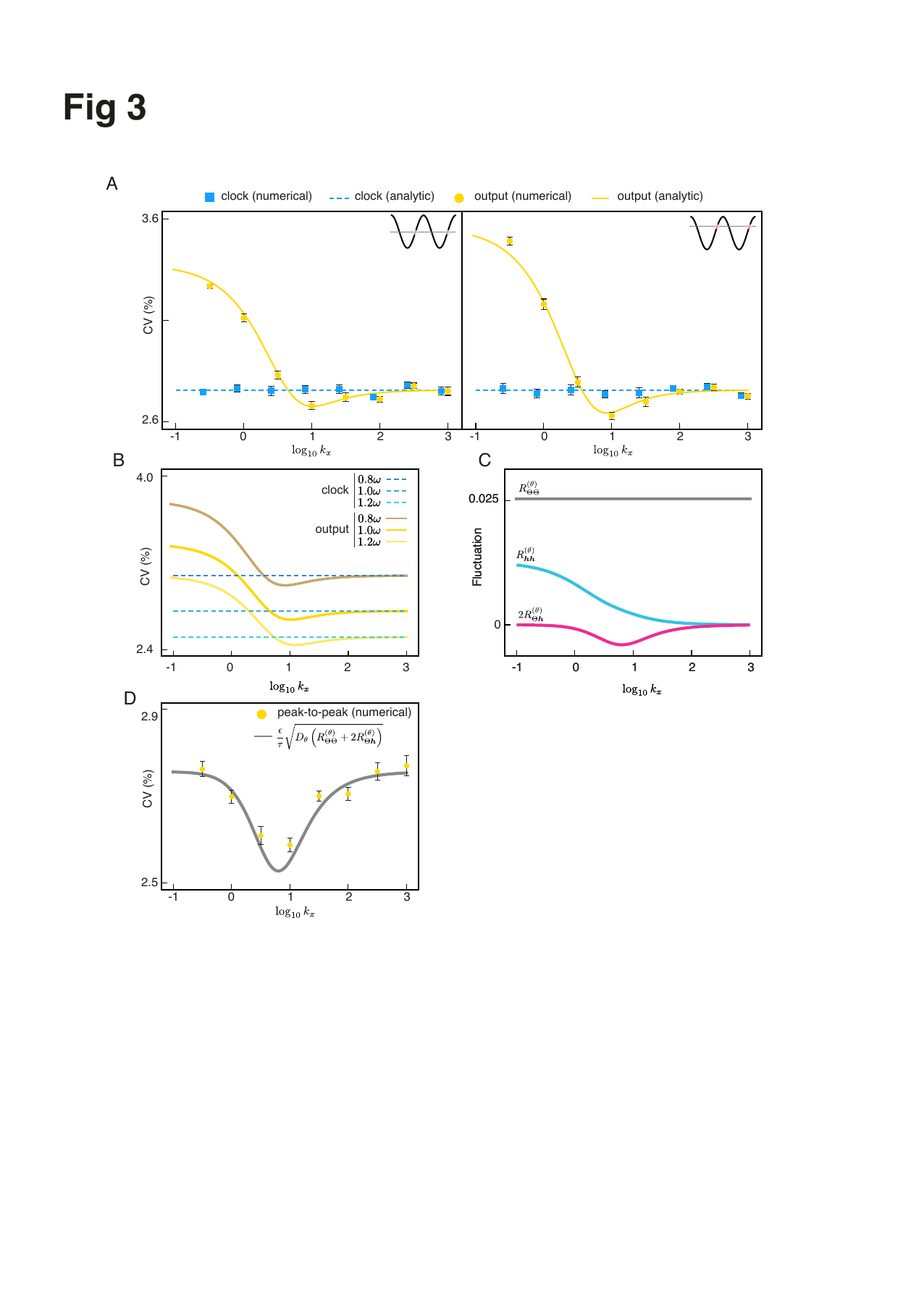}
\includepdf[pages=-]{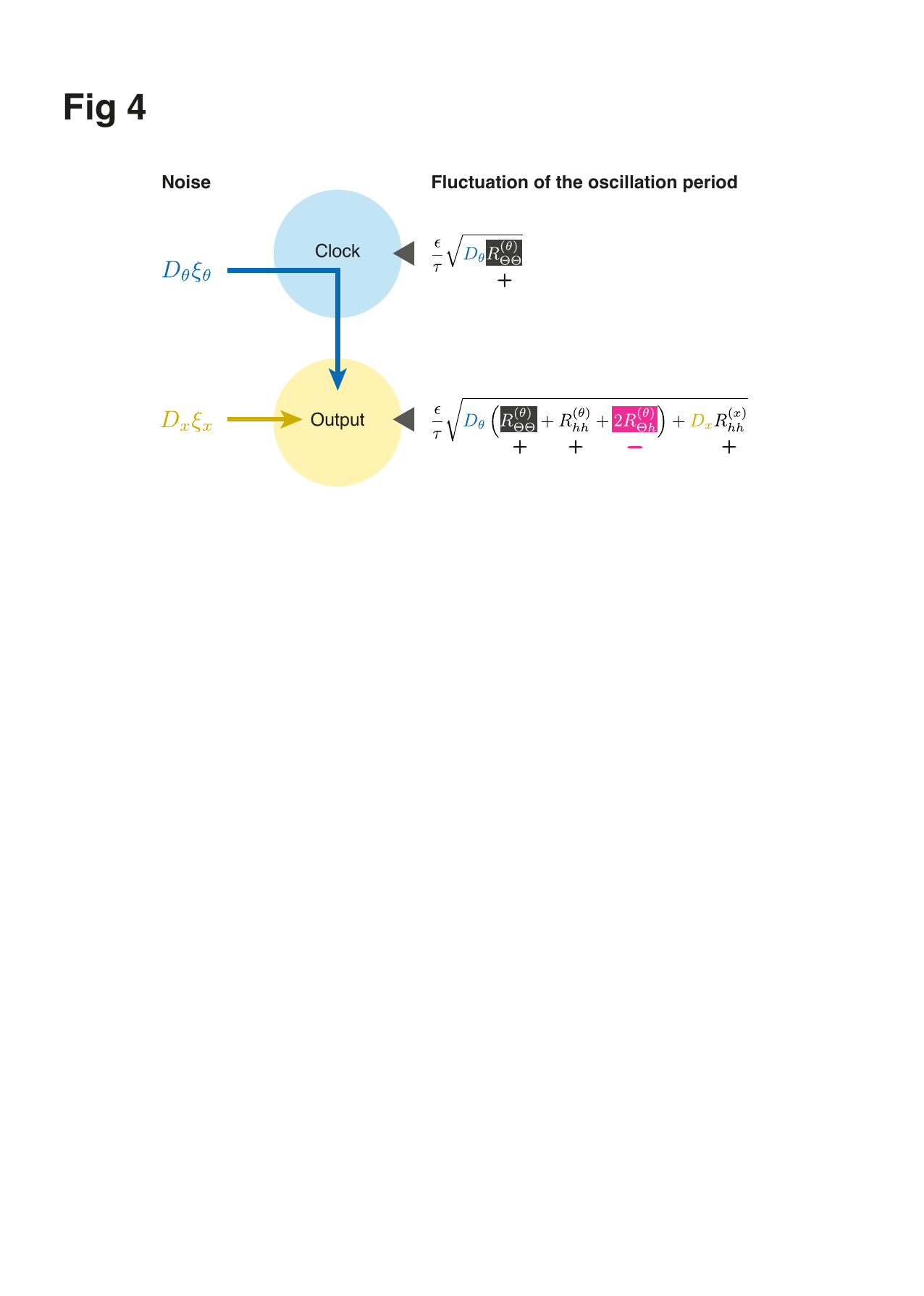}
\end{document}